\documentstyle[12pt]{article}
\begin{document}

\tolerance=5000

\def\be{\begin{equation}}
\def\ee{\end{equation}}
\def\bea{\begin{eqnarray}}
\def\eea{\end{eqnarray}}
\def\nn{\nonumber \\}
\def\e{{\rm e}}
\def\sqg{\sqrt{-g}}
\def\tr{{\rm tr}}

\  \hfill 
\begin{minipage}{3.5cm}
NDA-FP-30 \\
December 1996 \\
hep-th/yymmxxx \\
\end{minipage}

\ 

\vfill

\begin{center}

{\Large\bf
Effective Lagrangian and static black holes 
in 2D dilatonic gravity inspired by quantum effects}

\vfill

{\Large\sc Shin'ichi NOJIRI}\footnote{
e-mail : nojiri@cc.nda.ac.jp}
and
{\Large\sc Sergei D. ODINTSOV$^{\spadesuit}$}\footnote{
e-mail : odintsov@quantum.univalle.edu.co}

\vfill

{\large\sl Department of Mathematics and Physics \\
National Defence Academy \\
Hashirimizu Yokosuka 239, JAPAN}

{\large\sl $\spadesuit$ Dep.de Fisica \\
Universidad del Valle \\
AA25360, Cali, COLUMBIA \\
and \\
Tomsk Pedagogical Univer. \\
634041 Tomsk, RUSSIA}

\vfill

{\bf ABSTRACT}
\end{center}

We study the effective action in 2D dilaton-Maxwell 
quantum gravity.
Working with the one-loop renormalizable subset of such 
theories, we construct the improved effective Lagrangian 
which contains curvature under logarithm. This effective Lagrangian
leads to new classical dilatonic gravity inspired by
quantum effects. 
The static black holes (BH) solutions which may play 
the role of a remnant after the Hawking radiation for 
such theory are carefully investigated. 
The effective Lagrangian for Gross-Neveu-dilaton gravity 
is also constructed (in 1/N expansion).

\vfill

\noindent
PACS: 04.20.Gz, 11.25.Mj

\newpage

\section{Introduction}

There was recently much of interest in the study of 2D dilatonic gravity 
and its BH solutions 
(see refs.\cite{OS}--\cite{no}).

At the beginning of their study it has been often claimed 
that such models provide solvable toy models of quantum 
gravity and the Hawking radiation.
Unfortunately, it has become nowdays clear that 2D 
dilatonic gravity is not much easier than corresponding 
higher dimensional theories.

Usual Lagrangians for 2D dilatonic gravities are stringy-motivated theories.
It often happens that their Lagrangians are taken 
simply by hands. 
In such circumstances, it could be of interest to study 
some classes of dilatonic gravities which are selected by 
some principle. 
In this work, we suggest to start from the 
(one-loop) renormalizable subset
of 2D dilatonic gravity which induces the effective 
Lagrangian on quantum level. This effective Lagrangian
maybe the started point for the new model of dilatonic
gravity inspired by quantum effects and renormalizability.
It is quite reasonable as we are interesting generally 
speaking in renormalizable quantum gravity 
(forgetting about string paradigm).
Then we study static BH solutions for such theory.

At the next section, we work with general version of 
2D dilaton-Mawxwell gravity and find the improved 
effective Lagragian and static BH solutions for corresponding inspired
theory.
Section 3 is devoted to the discussion of the same question 
in 2D dilaton gravity interacting with Gross-Neveu model.
Here $1/N$ expansion scheme is adopted.

\section{Effective Lagrangian and static BH}

We now start with the general version of 
two-dimensional dilaton-Maxwell gravity with the action
\be
\label{gS}
S=\int d^2x \sqg \left[{1 \over 2}Z(\varphi)
g^{\mu\nu}\partial_\mu\varphi\partial_\nu\varphi
+C(\varphi)R+V(\varphi)+{1 \over 4}f_1(\varphi)F^2_{\mu\nu}
\right]
\ee
where $\varphi$ is dilaton, $Z(\varphi)$, $C(\varphi)$, 
$V(\varphi)$, $f_1(\varphi)$ are dilaton-depending functions. 
The theory with the action (\ref{gS}) (without 
electro-magnetic sector) has been introduced in refs.\cite{OS} and 
\cite{BO}
where its one-loop structure has been found.
In addition, the question of quantum equivalence of 
different models from class (\ref{gS}) as well as gauge dependence of 
effective action (EA) or calculation of gauge-fixing independent EA 
have been studied in \cite{ENO}.

We will be interesting here in the one-loop renormalizable subset 
of the theory (\ref{gS}). As it has been shown in refs.\cite{ENO} 
one can make the one-loop renormalization transformation of 
the metric
\be
\label{rgmtrc}
g_{\mu\nu}=\exp\left\{{1 \over \epsilon}\left[
{1 \over C(\varphi)} + {Z(\varphi) \over 2 (C'(\varphi))^2}
\right]\right\} g_{\mu\nu}^R
\ee
As we see only trace part of metric should be renormalized in 
two-dimensional dilaton-Maxwell gravity.
In this case the renormalized effective action is given as \cite{ENO} 
(the Lorentz gauge condition is used)
\bea
\label{SR}
S_R&=&\int d^2x \sqg \Bigl\{ {Z(\varphi) \over 2}
g^{\mu\nu}\partial_\mu\varphi\partial_\nu\varphi
+C(\varphi)R+V(\varphi)+{1 \over 4}f_1(\varphi)F^2_{\mu\nu} \nn
&& + {1 \over \epsilon}\left(-{V'(\varphi) \over C'(\varphi)}
+{V(\varphi)Z(\varphi) \over 2 (C'(\varphi))^2}\right)
+{F^2_{\mu\nu} \over 4\epsilon}\left(
-{f'_1(\varphi) \over C'(\varphi)}-{f_1(\varphi) Z(\varphi)
\over 2 (C'(\varphi))^2}\right)\Bigr\}
\eea
where $\epsilon=2\pi(n-2)$.
The easy check shows that the theory under consideration 
will be one-loop renormalizable in usual sense for the following 
potentials:
\bea
\label{renpot}
V(\varphi)&=&\exp\left\{aC(\varphi)
+\int{d\varphi\, Z(\varphi) \over 2C'(\varphi)} \right\} \ , \nn
f_1(\varphi)&=&\exp\left\{-bC(\varphi)
-\int{d\varphi\, Z(\varphi) \over 2C'(\varphi)} \right\}
\eea
where $a$ and $b$ are the arbitrary constants.

One interesting variant of the model (\ref{gS}) is given by
the choice of $Z(\varphi)=0$. 
In this case
\be
\label{rgmtrc2}
g_{\mu\nu}=\exp\left\{{1 \over \epsilon C(\varphi)}\right\} 
g_{\mu\nu}^R
\ee
and
\bea
\label{SR2}
S_R&=&\int d^2x \sqg \Bigl\{ 
C(\varphi)R+V(\varphi)+{1 \over 4}f_1(\varphi)F^2_{\mu\nu} \nn
&& + {1 \over \epsilon}\left(-{V'(\varphi) \over C'(\varphi)}\right)
+{F^2_{\mu\nu} \over 4\epsilon}\left(
-{f'_1(\varphi) \over C'(\varphi)}\right)\Bigr\}
\eea
The set of renormalizable potentials is given as following
\be
\label{rnpot}
V=\exp(aC(\varphi)) ,\ \ 
f_1(\varphi)=\exp(-bC(\varphi))
\ee
%One can also consider the theory with $Z(\varphi)=1$ theory as 
%another explicit example.

At the next step, we will write the explicit expressions for 
effective Lagrangian in above two cases. 
Note that because only trace part of metric tensor is 
transformed under renormalization (\ref{rgmtrc}), one 
can show that corresponding $\beta$-function of metric tensor is 
zero (see \cite{OS}, for example). 

Using the fact that theory is one-loop renormalizable one 
can apply the standard renormalization group considerations 
in curved space-time (see, for introduction \cite{BOS}),  
one can find the effective Lagrangian as the following (in the 
background field method, we consider the background fields 
to be: $\varphi=$almost static, $R=$const, $F_{\mu\nu}\sim$ const)
\bea
\label{EA}
L_{\rm eff}&=& {Z(\varphi) \over 2} 
g^{\mu\nu}\partial_\mu\varphi\partial_\nu\varphi
+C(\varphi)R+V(\varphi)- {1 \over 2}\left(-{V'(\varphi) \over C'(\varphi)}
+{V(\varphi)Z(\varphi) \over 2 (C'(\varphi))^2}\right)t \nn
&&+{1 \over 4}F^2_{\mu\nu}\left[f_1(\varphi)
+{1 \over 2}t\left({f'_1(\varphi) \over C'(\varphi)}
+{f_1(\varphi) Z(\varphi)
\over 2 (C'(\varphi))^2}\right)\right]\Bigr\}
\eea
where we suppose that $C(\varphi)R$ is biggest from background 
fields and dilatonic functions, then $t\sim\ln {C(\varphi)R \over \mu^2}$ ($C(\varphi)R$ under $\ln$ should be understood to be the 
absolute value of it when it is negative.). 
Note that in effective Lagrangian (\ref{EA}) we actually take 
into account the leading logarithm of perturbation serie
like in \cite{CW} , hence it is non-perturbative quantity.

Now, we consider new classical theory of dilatonic gravity which 
is inspired by the effective Lagrangian (8). Unlike the effective
Lagrangian (8) which was obtained for almost constant values of fields
and supposing some relations between fields, the new model is considered
to be the arbitrary classical theory without any relations between the fields.
 The Lagrangian is

\be
\label{EAb}
L_{\rm eff} = {Z(\varphi) \over 2} 
g^{\mu\nu}\partial_\mu\varphi\partial_\nu\varphi
+C(\varphi)R+\tilde V(\varphi)+{1 \over 4}F^2_{\mu\nu}\tilde f_1(\varphi)
\ee
where
\bea
\label{effpot}
\tilde V(\varphi)&=&V(\varphi)- {1 \over 2}\left(-{V'(\varphi) \over C'(\varphi)}
+{V(\varphi)Z(\varphi) \over 2 (C'(\varphi))^2}\right)
\ln {C(\varphi)R \over \mu^2} \\
\tilde f_1(\varphi)&=& f_1(\varphi)
+{1 \over 2}\left({f'_1(\varphi) \over C'(\varphi)}
+{f_1(\varphi) Z(\varphi)
\over 2 (C'(\varphi))^2}\right)\ln {C(\varphi)R \over \mu^2}
\eea

In the following, we consider the case for simplicity
\be
\label{smpl}
Z(\varphi)=0\ ,\ \ C(\varphi)=\varphi\ ,\ \ f_1(\varphi)=0
\ee
Then the action is given by
\be
\label{ea}
S=\int d^2x\sqrt{-g}\left\{\varphi R + m^2\e^{a\varphi}\left(
1+{a \over 2}\ln{\varphi R \over \mu^2}\right)\right\}\ .
\ee
This will be starting point for the study of static black holes 
in such a theory which we consider as classical gravity theory.

In the conformal gauge
\be
\label{cg}
g_{\pm\mp}=-{1 \over 2}\e^{2\rho}\ , \hskip 1cm g_{\pm\pm}=0\ ,
\ee
the equations of motion are given by
\bea
\label{vphi}
0&=&R+m^2a\e^{a\varphi}\left(1+{a \over 2}\ln{\varphi R \over \mu^2}
+{1 \over 2\varphi}\right) \ ,\\
\label{gpm}
0&=&\partial_+\partial_-\left( \varphi + m^2\e^{a\varphi}{a \over 2R}
\right)+{1 \over 4}m^2\e^{2\rho+a\varphi}\left(
1+{a \over 2}\ln{\varphi R \over \mu^2}-{a \over 2}\right) \ ,\\
\label{gpp}
0&=&\partial_\pm\left\{\e^{-2\rho}\partial_\pm\left(
\varphi + m^2\e^{a\varphi}{a \over 2R}\right)\right\}\ .
\eea
The first equation (\ref{vphi}) is obtained by the variation 
of $\varphi$, the second one (\ref{gpm}) by that of $g^{+-}$ 
and the third one (\ref{gpp}) by $g^{\pm\pm}$. 
Here we have used the fact that 
the variation of the scalar curvature, which has the 
following form in the conformal gauge 
\be
\label{R}
R=8\e^{-2\rho}\partial_+\partial_-\rho\ ,
\ee
is given by
\bea
\label{dR}
\delta R &=&\e^{-2\rho}\partial_+\partial_-\left\{\e^{2\rho}\left(
\delta g^{+-}+\delta g^{-+}\right)\right\}
-2\partial_+\partial_-\rho\left(
\delta g^{+-}+\delta g^{-+}\right) \nn
&&-\e^{-2\rho}\partial_+\left\{\e^{-2\rho}\partial_+\left(
\e^{4\rho}\delta g^{++}\right)\right\}
-\e^{-2\rho}\partial_-\left\{\e^{-2\rho}\partial_-\left(
\e^{4\rho}\delta g^{--}\right)\right\} .
\eea
Equation (\ref{gpp}) is integrated to be
\be
\label{vphi2}
\partial_\pm\left(
\varphi + m^2\e^{a\varphi}{a \over 2R}\right)
=\alpha^\pm (x^\mp)\e^{2\rho}\ .
\ee
Here $\alpha^\pm (x^\mp)$ are arbitrary functions.
Therefore we obtain
\bea
\label{vphi3}
&& \partial_+\partial_-\left(
\varphi + m^2\e^{a\varphi}{a \over 2R}\right)
 =\left( (\alpha^\pm)' (x^\mp)+\alpha^\pm (x^\mp)\partial_\mp\rho
\right)\e^{2\rho} \ ,
\eea
\be
\label{vphi4}
(\alpha^+)' (x^-)+\alpha^+ (x^-)\partial_- \rho
=(\alpha^-)' (x^+)+\alpha^- (x^+)\partial_+ \rho \ .
\ee
By using the residual symmetry in the conformal gauge
\be
\label{resy}
\rho\rightarrow \rho + f^+(x^-) + f^- (x^+)
\ee 
(Here $f^\pm (x^\mp)$ are arbitrary functions.), 
we can choose a coordinate system where 
$(\alpha^+)' (x^-)=(\alpha^-)' (x^+)=0$ and we obtain 
\be
\label{rh}
(\alpha^+\partial_- -  \alpha^-\partial_+)\rho=0\ .
\ee
Here $\alpha^\pm$ are constants.
Equation (\ref{rh}) tells that $\rho$ is a function of 
the only one
coordinate $x=\alpha^+ x^- + \alpha x^-$. If $x$ is time(space)-like, 
we can identify $x$ with the time (space) coordinate $t$ ($r$) and 
the solution expresses the uniform (static) space-time.
Now since we are interested in static solution, we replace 
$\partial_\pm$ in Eqs.(\ref{vphi}), (\ref{gpm}) and (\ref{gpp}) 
by $\pm{1 \over 2}\partial_r$ and we obtain
\bea
\label{vphi22}
0&=&R+m^2a\e^{a\varphi}\left(1+{a \over 2}\ln{\varphi R \over \mu^2}
+{1 \over 2\varphi}\right) \\
\label{gpm2}
0&=&-\partial_r^2\left( \varphi + m^2\e^{a\varphi}{a \over 2R}
\right)+m^2\e^{2\rho+a\varphi}\left(
1+{a \over 2}\ln{\varphi R \over \mu^2}-{a \over 2}\right) \\
\label{gpp2}
0&=&\partial_r\left\{\e^{-2\rho}\partial_r\left(
\varphi + m^2\e^{a\varphi}{a \over 2R}\right)\right\}\ .
\eea
Here the scalar curvature $R$ is given by
\be
\label{R2}
R=-2\e^{-2\rho}\partial_r^2\rho\ .
\ee
Equation (\ref{gpp2}) can be integrated to be
\be
\label{gpp2b}
\partial_r\left(
\varphi + m^2\e^{a\varphi}{a \over 2R}\right)
=\alpha\e^{2\rho}
\ee
Here $\alpha$ is a constant. 
By substituting (\ref{gpp2b}) into (\ref{gpm2}), we 
obtain
\be
\label{gpm2b}
0=-2\alpha\partial_r\rho+m^2\e^{2\rho+a\varphi}\left(
1+{a \over 2}\ln{\varphi R \over \mu^2}-{a \over 2}
\right) \ .
\ee
The special solution where 
\be
\label{alph}
\alpha=0
\ee
can be found. When $\alpha=0$, from the 
equation (\ref{gpm2b}) we obtain 
\be
\label{gpm2c}
R=-{\mu^2 \over \varphi}\e^{1-{2 \over a}}
\ee
By substituting the equation (\ref{gpm2c}) 
into (\ref{vphi22}), we obtain
\be
\label{vphi33}
\e^{-a\varphi}={m^2a^2 \over 2\mu^2}\e^{{1 \over a}-1}
\varphi+{m^2a \over 2\mu^2}\e^{{1 \over a}-1}
\ee
This tells that $\varphi$ is a constant
\be
\label{vphi44}
\varphi=\varphi_0\ .
\ee
The solution $\varphi_0$ exists when
\bea
\label{sol}
& a>0 & \\
\mbox{or}\ & a<0 & \ \mbox{and}\ 
-{m^2 a \over 2\mu^2}\e^{{2 \over a}-3}\geq 1 \ .
\eea 
By using (\ref{gpm2c}), we find that the scalar 
curvature $R$ is a constant
\be
\label{R3}
R=R_0\equiv -{\mu^2 \over \varphi_0}\e^{1-{2 \over a}}\ .
\ee
%$-R_0$ (or $\varphi_0$) is positive when
%\bea
%\label{Rsign}
%& a>0 &\ {m^2 a \over 2\mu^2}\e^{{2 \over a}-1} < 1 \\
%\mbox{or}\ & a<0 &
%\eea
$\rho$ satisfies, due to Eqs. (\ref{R2}) and (\ref{R3}),
\be
\label{rho}
\partial_r^2 \rho =-{R_0 \over 2}\e^{2\rho}
\ee
Equation (\ref{rho}) is integrated to be
\be
\label{rho2}
(\partial_r \rho)^2=-{R_0 \over 2}\left(\e^{2\rho}+C
\right)
\ee
Here $C$ is a constant.
The solutions of Equation (\ref{rho2}) are given by 
\be
\label{solt}
\e^{2\rho}={C \over 4}
\left(\cot {\sqrt{CR_0 \over 8}r} 
+  \tan {\sqrt{CR_0 \over 8}r} \right)^2
\ee
(When $CR_0<0$, 
$\tan {\sqrt{CR_0 \over 8}r}
=i\tanh {\sqrt{|CR_0| \over 8}r}$.)
In the limit $C\rightarrow 0$, we obtain
\be
\label{C0}
\e^{2\rho}\rightarrow {4 \over R_0 r^2}
\ee
If we define new coordinates $\zeta^\pm$ 
when $CR_0<0$ by
\be
\label{zet}
\zeta^\pm =\e^{\sqrt{-{CR_0 \over 8}}(r\pm t)}
\ee
the metric tensor is given by
\be
\label{metric}
ds^2={8 \over R_0}{1 \over (\zeta^+\zeta^-)^2-1}d\zeta^+d\zeta^-
\ee
If we change the role of the time coordinate $t$ and the 
space coordinate $r$, we obtain the uniform solution and 
the metric (\ref{metric}) expresses the de Sitter or the 
anti-de Sitter universe with the constant curvature.
The temperature of the obtained static object vanishes 
since $|\e^{2\rho}|\geq |C|$ in Equation (\ref{solt}) 
and $\e^{2\rho}$ cannot vanish. Therefore the static solution 
could be a remnant after the Hawking radiation.

In the following, we consider the Maxwell-dilaton gravity, 
as another interesting model, with\footnote{
The Maxwell-dilaton gravity with 
\[
Z(\varphi)={C(\varphi) \over 4}=\e^{-2\varphi}\ , 
\ \ f_1(\varphi)=\e^{2\varphi}
\]
is classically solvable and the quantum effect
was studied in Refs.\cite{no}. } 
\be
\label{smpl2}
Z(\varphi)=0\ ,\ \ C(\varphi)=\varphi\ ,
\ee
and
\be
\label{smpl2b}
V(\varphi)=m^2\e^{a\varphi}\ , \ \ 
f_1(\varphi)=\e^{-b\varphi}\ .
\ee
We now assume that $\e^{-b\varphi}F^2_{\mu\nu}$ is the biggest 
background field, therefore
\be
\label{ttt}
t=\ln{f_1(\varphi)F^2_{\mu\nu} \over \mu^2}\ .
\ee
($f_1(\varphi)F^2_{\mu\nu}$ under $\ln$ should be understood 
to be the 
absolute value of it when it is negative.)
Then the action (in the same sense as (9)) is given by
\bea
\label{eael}
S&=&\int d^2x\sqrt{-g}\Bigl\{\varphi R + m^2\e^{a\varphi}\left(
1+{a \over 2}\ln{\e^{-b\varphi}F^2_{\mu\nu} \over \mu^2}
\right) \nn
&&+{1 \over 4}\e^{-b\varphi}F^2_{\mu\nu}\left(
1-{b \over 2}\ln{\e^{-b\varphi}F^2_{\mu\nu} \over \mu^2}
\right)
\Bigr\}\ ,
\eea
and the equations of motion are given by
\bea
\label{vphiel}
0&=&R+m^2a\e^{a\varphi}\left(1+{a \over 2}
\ln{\e^{-b\varphi}F^2_{\mu\nu} \over \mu^2}
-{b \over 2}\right) \nn
&&-{b \over 4}\e^{-b\varphi}F^2_{\mu\nu}\left(
1-{b \over 2}\ln{\e^{-b\varphi}F^2_{\mu\nu} \over \mu^2}
-{b \over 2}\right) \ ,\\
\label{gpmel}
0&=&\partial_+\partial_- \varphi 
+{1 \over 4}m^2\e^{2\rho+a\varphi}\left(
1+{a \over 2}\ln{\e^{-b\varphi}F^2_{\mu\nu} \over \mu^2}
- a \right) \nn
&& -{b \over 16}\e^{2\rho-b\varphi}F^2_{\mu\nu}\left(
1-{b \over 2}\ln{\e^{-b\varphi}F^2_{\mu\nu} \over \mu^2}
- b\right) \ ,\\
\label{gppel}
0&=&\partial_\pm\left\{\e^{-2\rho}\partial_\pm \varphi\right\}\ ,
\\
\label{A}
0&=&\partial_\pm\left[\left\{
{1 \over 4}\e^{-b\varphi}\left(
1-{b \over 2}\ln{\e^{-b\varphi}F^2_{\mu\nu} \over \mu^2}
- {b \over 2}\right) + {1 \over 2}m^2a\e^{a\varphi}
{1 \over F^2_{\mu\nu}}
\right\}\sqrt{-F^2_{\mu\nu}}
\right]\ .
\eea
Here
\be
\label{F}
F^2_{\mu\nu}=-8\e^{-4\rho}\left(\partial_+ A_- - 
\partial_- A_+ \right)^2
\ee
The last equation (\ref{A}) is given by the variation of the 
gauge fields $A_\pm$ and  tells that 
\bea
\label{AA}
&& \left\{
{1 \over 4}\e^{-b\varphi}\left(
1-{b \over 2}\ln{\e^{-b\varphi}F^2_{\mu\nu} \over \mu^2}
- {b \over 2}\right) + {1 \over 2}m^2a\e^{a\varphi}
{1 \over F^2_{\mu\nu}}
\right\}\sqrt{-F^2_{\mu\nu}}\nn
&&=\beta\ \mbox{(constant)}\ .
\eea
By the argument similar to the previous one in Eqs.(\ref{vphi2}) 
$\sim$ (\ref{rh}), Eq.(\ref{gppel}) 
tells the solution is static or uniform. When we assume that 
the solution is static, the equations of motion has the following 
form;
\bea
\label{vphiel2}
0&=&-2\partial_r^2 \rho + m^2a\e^{2\rho+a\varphi}\left(1+{a \over 2}
\ln{\e^{-b\varphi}F^2_{\mu\nu} \over \mu^2}
-{b \over 2}\right) \nn
&&-{b \over 4}\e^{2\rho-b\varphi}F^2_{\mu\nu}\left(
1-{b \over 2}\ln{\e^{-b\varphi}F^2_{\mu\nu} \over \mu^2}
-{b \over 2}\right) \ ,\\
\label{gpmel2}
0&=&-\partial_r^2 \varphi 
+m^2\e^{2\rho+a\varphi}\left(
1+{a \over 2}\ln{\e^{-b\varphi}F^2_{\mu\nu} \over \mu^2}
- a \right) \nn
&&-{b \over 4}\e^{2\rho-b\varphi}F^2_{\mu\nu}\left(
1-{b \over 2}\ln{\e^{-b\varphi}F^2_{\mu\nu} \over \mu^2}
- b\right) \ ,\\
\label{gppel2}
0&=&\partial_r\left\{\e^{-2\rho}\partial_r \varphi\right\}\ .
\\
\label{A2}
0&=&\partial_r \left[\left\{
{1 \over 4}\e^{-b\varphi}\left(
1-{b \over 2}\ln{\e^{-b\varphi}F^2_{\mu\nu} \over \mu^2}
- {b \over 2}\right) + {1 \over 2}m^2a\e^{a\varphi}
{1 \over F^2_{\mu\nu}}
\right\}\sqrt{-F^2_{\mu\nu}}
\right]
\eea
and
\be
\label{F2}
F^2_{\mu\nu}=-{1 \over 2}\e^{-4\rho}(\partial_r A)^2
\ee
Here
\be
\label{AAA}
A\equiv A_t={A_+ + A_- \over 2}
\ee
Equation (\ref{gppel2}) can be integrated to be
\be
\label{pvphi}
\partial_r\varphi=\alpha\e^{2\rho}\ .
\ee
($\alpha$ is a constant.)
Therefore if
\be
\label{calpha}
\alpha=0
\ee
$\varphi$ is a constant and $F^2_{\mu\nu}$ is also a constant
due to Equation (\ref{AA}).
In the following, we consider the following case
\be
\label{simpl}
\alpha=\beta=0\ ,\ \ \ a=b=1
\ee
for simplicity.
Then by using Equations (\ref{AA}) and (\ref{gpmel2}), 
we find 
\bea
\label{solel}
\varphi &=& -\ln\left({4m^2 \over \mu^2}\right) \ , \nn
F^2_{\mu\nu} &=& -{\mu^4 \over 4m^2}\ , \nn
\Bigl(A&=&{\mu^2 \over m\sqrt 2 }\int^r dr \e^{2\rho}\Bigr).
\eea
By substituting (\ref{solel}) into (\ref{vphiel2}), 
we find
\be
\label{rhoel}
0=\partial_r^2\rho+{\mu^2 \over 8}\e^{2\rho}\ .
\ee
Therefore the scalar curvature is a constant
\be
\label{curv}
R={\mu^2 \over 2}
\ee
The solution for the constant curvature is given in Eq.(\ref{solt}).
For general $\beta$, $a$ and $b$, the qualitive nature of 
the solution does not change.

\section{Effective Lagrangian in dilaton gravity with Gross-Neveu 
model}

Let us consider now Gross-Neveu model \cite{GN} interacting with dilaton gravity (\ref{ea})
\bea
\label{eaGN}
S&=&\int d^2x\sqrt{-g}\left\{\varphi R + m^2\e^{a\varphi}\left(
1+{a \over 2}\ln{\varphi R \over \mu^2}\right)\right.\nn
&&\left. +i\bar\psi\gamma^\mu D_\mu\psi +\bar\psi\psi\sigma
-{\sigma^2 N \over 2\lambda}\right\}\ .
\eea
where $\lambda$ is the four-fermion coupling constant, $\psi$ is 
$N$-component spinor and $\sigma$ is the auxiliary field.
It is quite well-known that theory under discussion, is asymptotically 
free in $1/N$-expansion in flat space-time.
Note also that one can add to (\ref{eaGN}) Maxwell term without 
any problems. However, it doesn't interact explicitly with fermions, 
hence it actually doesn't influence the dynamics of the theory.

We are going to discuss the theory with the action (1) in semi-classical 
approach, working in $1/N$-expansion and supposing that curvature is 
constant.

Then we should add to the action (\ref{eaGN}) conformal anomaly 
term \cite{CGHS,no}
\be
\label{cnfanly}
S_c=\int d^2x\sqrt{-g}\left\{-{1 \over 2}(\nabla Z)^2
+\sqrt{N \over 48} ZR \right\}
\ee
where $Z$ is the auxiliary scalar field. 
Also, in the leading order of $1/N$-expansion, dilatonic quantum effects 
are subleading.
The leading term which should be added to (\ref{eaGN}) is given by 
Gross-Neveu effective potential in constant positive curvature 
space-time \cite{IMO};
\bea
\label{dSea}
S_V&=&-N\int d^2x\sqrt{-g}\left\{\sigma^2\left[{1 \over 2\lambda}
-{\tr 1 \over 8\pi}\left(2 + \ln {2\mu^2 \over R}\right)\right]\right.\nn
&&\left. 
+{\tr 1 \over 4\pi}\int_0^\sigma ds\, s\left[\Psi(1+is\sqrt{2 \over R})
+\Psi(1-is\sqrt{2 \over R})\right]
\right\}
\eea
where $\tr 1=2$, $\Psi(z)$ is logarithmic derivative of 
$\Gamma$-function.
Working with the action 
\be
\label{stot}
S_{{\rm tot}}=S+S_c+S_V
\ee
in the same way as it has been done in \cite{MOS}, we 
may again find static BH solutions of the same sort as above.
Their qualitative structure is the same as above, 
so we do not discuss them in detail.

It is very interesting to note that one can make 
renormalization group (RG) improvement of $S_V$ (\ref{dSea})
due to renormalizability and asymptotic freedom.
One has only change the coupling constant ($\lambda=Ng^2$) 
in the effective action by the running coupling constant;
\be
\label{ren}
g^2(t)={g^2 \over 1-{\lambda t \over \pi}}
\ee
where $t$ is RG parameter. Working in the regime of 
strong curvature, $t=\ln{R \over \mu^2}$.
Hence, asymptotic freedom in Gross-Neveu model is induced by BH 
curvature.
Stronger BH curvature induces stronger asymptotic freedom as in 
$d=4$ gauge theories in curved spacetime \cite{BOS}.
In other words, in the vicinity of constant curvature 
BH, Gross-Neveu model tends to become free theory. 
This effect (inducing of asymptotic freedom by BH curvature) 
may have the important cosmological applications.

Note finally that it it not difficult to generalize above 
construction for asymptotically free supersymmetric theories.
One can consider conformal theories with dilatonic supergravity 
\cite{SBH}
or supergauge theories with dilatonic supergravity.
One may expect that BH solutions in such theories may not only lead to  induced asymptotic freedom but also to 
dynamical supersymmetry breaking.

\noindent
{\bf Acknowledgement}

SDO would like to thank COLCIENCIAS(COLOMBIA) for a partial support 
of this work.

\end{document}